\documentclass[twocolumn,superscriptaddress,nofootinbib]{revtex4}

\usepackage{amssymb}
\usepackage{amsmath}
\usepackage{multirow}
\usepackage{mathrsfs}
\usepackage{graphicx}
\usepackage{color}
\usepackage{dcolumn}
\usepackage{bm}
\newfont{\Fr}{eufm10}
\usepackage{graphics}
\usepackage{epsfig}
\usepackage[english]{babel}
\usepackage{verbatim}
\usepackage{amsfonts}
\usepackage[latin1]{inputenc}
\usepackage{hyperref}
\RequirePackage[english]{babel}
\RequirePackage{amsmath} % Useful for equation alignment
\RequirePackage{verbatim} % Allows to have comments over multiple lines
\RequirePackage{mathrsfs}
\RequirePackage{amsfonts}
\RequirePackage{amssymb}
\RequirePackage[latin1]{inputenc}
\RequirePackage{hyperref}

\newcommand{\beq}{\begin{equation}}
\newcommand{\eeq}{\end{equation}}
\newcommand{\barr}{\begin{eqnarray}}
\newcommand{\earr}{\end{eqnarray}}
\newcommand{\nk}{\textbf{k}}
\newcommand{\dphi}{\delta \varphi}
\newcommand{\x}{\textbf{x}}

\newcommand{\bra}{\langle}
\newcommand{\ket}{\rangle}
\newcommand{\mH}{\mathcal{H}}
\newcommand{\mP}{\mathcal{P}}

\newcommand{\nx}{\vec{x}}

\newcommand{\nn}{\nonumber \\}

\begin{document}

\title{Emergence of inflationary perturbations in the CSL model}
\author{Gabriel Le\'{o}n}
\email{gleon@df.uba.ar} \affiliation{Departamento de
F\'\i sica, Facultad de Ciencias Exactas y Naturales, Universidad
de Buenos Aires, Ciudad Universitaria - Pab.I, 1428 Buenos Aires,
Argentina}
\author{Gabriel R. Bengochea}
\email{gabriel@iafe.uba.ar} \affiliation{Instituto de Astronom\'\i
a y F\'\i sica del Espacio (IAFE), UBA-CONICET, CC 67, Suc. 28,
1428 Buenos Aires, Argentina}

\begin{abstract}

The inflationary paradigm is the most successful model that explains
the observed spectrum of primordial perturbations. However, the precise emergence of such
inhomogeneities and the quantum-to-classical transition of the perturbations has not yet reached a
consensus among the community. The Continuous Spontaneous Localization model
(CSL), in the cosmological context, might be used to provide a solution
to the mentioned issues by considering a dynamical reduction of the wave
function. The CSL model has been applied to the inflationary
universe before and different conclusions have been obtained. In this
letter, we use a different approach to implement the CSL model during inflation. In
particular, in addition to accounting for the quantum-to-classical transition, we use the
CSL model to generate the primordial perturbations, that
is, the dynamical evolution provided by the CSL model is responsible for the
transition from a homogeneous and isotropic initial state to a final one lacking such
symmetries. Our approach leads to results that can be clearly distinguished
from preceding works. Specifically, the scalar and tensor power
spectra are not time-dependent, and retains the amplification mechanism of the CSL model.
Moreover, our framework depends
only on one parameter (the CSL parameter) and its value is consistent with cosmological
and laboratory observations.

\end{abstract}

\pacs{Valid PACS appear here}


\maketitle

\section{Introduction}\label{intro}

The inflationary paradigm is held among the majority of cosmologists
as a successful model for addressing the primordial inhomogeneities that
represent the seeds of cosmic structure. In fact, recent observations
from the Cosmic Microwave Background (CMB) radiation
\cite{wmap9cosmo,Planckcls13,Planck13inflation,PlanckBicep15,planckcmb2,Planckinflation15}
are quite consistent with
the standard prediction from the simplest inflationary model, namely, the
prediction of a nearly scale-invariant power spectrum. On the other hand, the
exact physical mechanism responsible for the generation of the primordial
curvature perturbations, associated to a highly Gaussian stochastic classical
field is still a matter of debate. In
particular, inflation is based on a combination of quantum mechanics and
general relativity, two theories that are difficult to merge at both the
conceptual and technical level. Therefore, it is expected that when these
theories are used in the same footing, some difficulties might arise.
Specifically, the quantum-to-classical transition, a subject which has been
present since the conception of the quantum theory, is an issue that has not
been fully resolved in the early inflationary universe.

A more precise formulation of the problem at hand can be stated as follows: the
inflationary universe is characterized by a background spacetime that is
completely homogeneous and isotropic. Equivalently, the quantum state of the
matter field, namely the vacuum state, is also perfectly homogeneous and
isotropic (i.e. it is an eigenstate of the operators associated to the generators of
spatial translations and rotations). On the other hand, the present universe is the result
of the evolution of primordial density inhomogeneities.

Henceforth, the problem is: how
do the inhomogeneities and anisotropies originate from the initial
highly symmetric (the symmetry being the homogeneity and isotropy) state of
the universe, described by both the background spacetime and the vacuum state
of the matter fields, given that the dynamical evolution, provided by
Schr\"odinger and Einstein equations, does not break the translational and
rotational symmetries? In other words, if one considers quantum mechanics as a
fundamental theory applicable, in particular, to the universe as a whole, then one
must regard any classical description of the state of any system as a sort of imprecise
characterization of a complicated quantum-mechanical state. The
universe we observe today is clearly well described by an inhomogeneous and
anisotropic classical state; therefore, such description must be considered as
an imperfect description of an equally inhomogeneous and anisotropic quantum
state. Consequently, if we want to consider the  inflationary  account  as
providing the physical  mechanism  for the generation of  the seeds of
structure, such account must contain an explanation for why the quantum state
that describes our actual universe does not possess the same symmetries  as  the
early quantum state of the universe, which happened to be perfectly symmetric. Since there
is nothing in the dynamical evolution (as given by the standard inflationary approach) of
the quantum state that can break those symmetries, the traditional inflationary
paradigm is incomplete in that sense (sometimes, an usual argument is that since the
vacuum fluctuations are not zero, somehow the system contains inhomogeneities. See Appendix \ref{appA}
for a discussion). Here, we also refer the reader to Refs.
\cite{zeh,sudarskySSB} where the relation between symmetry-breaking vacuum
and the reduction of the wave function is discussed.

Earlier works based on decoherence \cite{kiefer,halliwell,kiefer2,polarski} and the
consideration that the initial vacuum state of the universe evolves into a
highly squeezed state \cite{grishchuk}, led to a partial understanding of the
issue, namely, that the predictions from the quantum theory are
indistinguishable from those of a theory in which the random fluctuations are
the result of a classical stochastic process \cite{martin2007}. Nevertheless,
this argument by itself cannot address the fact that a single (classical)
outcome emerges from the quantum theory. In other words, decoherence (and the
squeezing of quantum states) cannot solve the quantum measurement
problem \cite{adler,schlosshauer}, a complication that, within the cosmological
context, is amplified due to the impossibility of recurring to the ``for all
practical purposes'' argument in the familiar laboratory situation; i.e. it is not
clear how to define in the primordial universe entities such as observers, detectors, etc.
Other cosmologists seem to adopt the Everett ``many-worlds'' interpretation of
quantum mechanics plus the decoherence process when confronted with the quantum-to-classical
transition in the inflationary universe \cite{mukhanov2005}. In the Everettian
formulation, reality is made of a connected weave of ever-splitting worlds, each one
realizing one of the alternatives that is opened by what we would call a
quantum-mechanical measurement. Regarding this point, we would like to refer the reader
to Refs. \cite{shortcomings,kent,stapp} where arguments against the Everett
interpretation are presented.

One possible way to address the mentioned issues or, in other words, to avoid the
standard ``measurement problem'' of quantum theory, is to invoke the
collapse of the wave function but without relying on any external objects, e.g.
\cite{PSS,shortcomings}. Other approaches to this problem have been based on
Bohmian quantum mechanics; see for instance \cite{pintoneto,valentini}.

The idea of invoking a self-induced collapse in order to generate the
primordial perturbations has been explored in great detail in previous
works, e.g.
\cite{adolfo2008,sigma,alberto,gabriel,susana2013,susana2014,finl,noninfl,bispec,pia}.
Moreover, in \cite{susana2012} some generic collapse schemes have been
tested using observational data coming from the 7-year release of the WMAP
collaboration \cite{wmap7data} and the Sloan Digital Sky Survey \cite{SDSS}.
Therefore, the attempt to solve the aforementioned issue is not just a matter
of philosophical concern but a relevant problem from the theoretical point
of view, yielding predictions that can be confronted with observational data.

Furthermore, the proposal of a self-induced collapse of the wave function has
been an active line of research since the early ideas of Di\'osi
\cite{diosi1987,diosi1989} and Penrose \cite{penrose1996}, advocating gravity as
the main
agent triggering the collapse. Also, the Ghirardi-Rimmini-Weber (GRW)
\cite{GRW} model was among the first attempts to introduce an objective
collapse model. In past decades, the Continuous Spontaneous Localization (CSL)
model, which can be viewed as a continuous version of the GRW model, has been
regarded as a promising model that can provide a solution to the quantum
measurement problem \cite{pearle,pearle2}. In particular, the CSL mechanism is
based on a non-linear stochastic modification of the standard Schr\"odinger
equation, in this way, spontaneous and random collapses of the wave function
occur all the time, to all particles, regardless they are isolated or interacting. The
idea
behind the CSL model \cite{bassi}, sometimes referred to as the
``amplification mechanism'', is that the collapses must be rare for microscopic
systems, in order not to alter their quantum behavior as described by the
Schr\"odinger equation. At the same time, their effect must increase when
several particles are hold together forming a macroscopic system. Moreover, the
testable predictions made by the CSL model are now considered to be feasible
within the current available technology \cite{mohammed}. On the other hand, the
CSL model is, at this stage, a non-relativistic model and therefore a complete
generalization to quantum fields is still under development \cite{tumulka,pearle3}.
Nevertheless, the particular features of the CSL model, e.g. the absence to
rely on external agents to solve the measurement problem,  would make it a viable
candidate to address the issue concerning the emergence of the classical primordial
perturbations if it predicts a spectrum consistent with the observational data.

In \cite{jmartin}, the CSL model was applied to the inflationary universe for the first
time. In this reference, the authors followed an approach
in which the collapse affected in the same way all modes of the
inflaton field and, as a consequence, the amplification mechanism of the CSL model
was lost. Additionally, their predicted scalar power spectrum contained some
features that conflicted with the nearly scale-invariant power spectrum, which
is consistent with the CMB data. In order to retain the amplification mechanism,
other authors \cite{hinduesS}, proposed a phenomenological manner in
which the CSL model would affect each mode of the inflaton field. Additionally, they
also assumed that the modification to Schr\"odinger equation, provided by the CSL
mechanism,
was by introducing a time-dependent parameter. This last assumption resulted in a
prediction for the scalar (and also tensor) power spectrum with the correct
shape (for a particular combination of their three free parameters) but that was
time-dependent. In order to overcome this shortcoming, the power spectrum was
chosen to be evaluated at the end on inflation. However, this choice makes their
CSL collapse parameter extremely small ($\sim e^{-120}$) \cite{hinduesT}.

A shared feature of the works in Refs. \cite{jmartin,hinduesS} is that the authors
worked in a joint metric-matter quantization of the perturbations characterized
by the Mukhanov-Sasaki variable \cite{mukv}. On the other hand, in Ref.
\cite{pedro} one followed a different approach to the problem; the authors
successfully applied the CSL collapse mechanism to the inflationary universe
but within the semiclassical gravity framework. Consequently, as pointed out
in \cite{finl,lucila}, the amplitude of the primordial gravitational waves is
exactly zero at first order in the perturbations. Therefore, a confirmed
detection of primordial gravity waves would make this approach face serious
issues.

Another distinction between the approaches in Refs. \cite{jmartin,hinduesS} and
\cite{pedro} is the specific role played by the self-induced collapse. This
difference is subtle but important. The authors of \cite{jmartin,hinduesS}
employed the CSL mechanism with the intention of
localizing the evolved inflaton vacuum state in an eigenstate of the
Mukhanov-Sasaki variable. Meanwhile, the authors of \cite{pedro} attributed
to the self-induced collapse the action of generating the primordial curvature
perturbation, irrespectively of which eigenstate the initial state evolves
into. Therefore, if no self-induced collapse occurs, the predicted power
spectrum in Refs. \cite{jmartin,hinduesS} is exactly the same as the standard
one. On the contrary, in Ref. \cite{pedro} the absence of a collapse results in
no perturbations of the spacetime at all, consequently, the predicted scalar power spectrum is exactly zero (see Sect. \ref{cslinflation} and Appendix \ref{appB} for a more detailed discussion).

In the present work, we adopt the same role for the collapse, provided
by the CSL model, as the one presented in \cite{pedro}, but we apply it to
the joint metric-matter quantization of the inflaton field, to improve the standard
inflationary treatment
providing a mechanism for the emergence of the primordial curvature perturbations, and
also for driving the quantum-to-classical transition.
In this way, we obtained predictions that are different from the ones in
\cite{jmartin,hinduesS}
and are also quite consistent with the
observational data. In particular, our predicted scalar and tensor power
spectra are not time-dependent, and the implementation of the CSL mechanism
retains the amplification mechanism. Also, our model introduces just one free parameter
and
in contrast with Ref. \cite{pedro}, it predicts a non-vanishing tensor power spectrum at
first order in the perturbations.

The article is organized as follows: in Sect. \ref{Secscalar} we review some basics
about the CSL collapse model applied to the inflationary universe, and we obtain the
scalar power spectrum within such a framework; in Sect. \ref{Sectensor} we show our
results for the power spectrum of tensor modes and the tensor-to-scalar ratio; in
Sects. \ref{discusion} and \ref{preworks} we make a discussion of our results and compare them
with previous works, and finally in Sect. \ref{conclusions} we summarize our
conclusions.

\section{The CSL model and the scalar power spectrum}
\label{Secscalar}

In this section, we will summarize some concepts regarding the CSL collapse
model, and we will use it to obtain a prediction for the primordial power
spectrum of scalar perturbations.

\subsection{Classical description of the perturbations}

The inflationary universe is described by Einstein equations $G_{ab} = 8 \pi G
T_{ab}$ ($c=1$) and the dynamics of the matter fields dominated by the
inflaton.
We will assume the simplest inflationary model, which is a single scalar field
$\varphi$ in the slow-roll approximation. The background spacetime is
accelerating in a quasi-de Sitter type of expansion, characterized by $\mH
\simeq -1/[\eta(1-\epsilon)]$, with $\mH \equiv a'/a$ the conformal expansion
rate, $a$ being the scale factor and the slow-roll parameter is defined as
$\epsilon \equiv 1-\mH'/\mH^2$ (a prime denotes partial derivative with respect
to conformal time $\eta$). The energy density of the universe is dominated by
the potential of the inflaton field $V$, and during the slow-roll inflation is satisfied
the condition $ \epsilon \simeq  M_P^2/2 (\partial_\phi V/V)^2 \ll 1$, with
$M_P^2 \equiv (8\pi G)^{-1}$ the reduced Planck mass. Since we will work in a
full quasi-de Sitter expansion, another useful parameter to characterize slow-roll
inflation is the second slow-roll parameter, i.e.
$\delta \equiv \epsilon - \epsilon'/2\mH \epsilon \ll 1$.

The scalar metric perturbations in a FRW background spacetime are generically
described by the line element:
\barr
ds^2 &=& a^2 (\eta) \big\{  -(1-2\phi) d\eta^2 + 2 (\partial_i B) dx^i d\eta
+\nonumber \\
&+& [ (1-2\psi) \delta_{ij} + 2 \partial_i \partial_j E] dx^i dx^j \big\}.
\earr

Since we will focus on a joint quantization of the metric and matter
perturbations, it is convenient to work with the gauge-invariant quantity known
as the Bardeen potential \cite{bardeen} defined as $\Phi \equiv \phi +
\frac{1}{a} [a (B-E')]'$. The matter sector dominated by the inflaton is
separated in an
homogeneous part plus small perturbations $\varphi (\eta,\x) = \varphi_0 (\eta)
+ \dphi (\eta,\x)$. In a similar way, the perturbations of the inflaton can be
modeled by the gauge-invariant fluctuation of the scalar field
$\dphi^{(\textrm{GI})} (\eta,\x) = \dphi + \phi_0' (B-E')$. The two objects
$\Phi$ and $\dphi^{(\textrm{GI})}$ can be used to form a new quantity called
the Mukhanov-Sasaki variable \cite{mukv} i.e.,
\beq\label{MS}
v(\eta,\x) \equiv a \left[ \dphi^{(\textrm{GI})} + \frac{\phi_0'}{\mH} \Phi
\right].
\eeq
Written in a gauge-invariant way, and, in the absence of anisotropic stress, the
components 00 and 0i of the Einstein perturbed equations at linear order
$\delta
G_{ab} = 8 \pi G \delta T_{ab}$, can be combined to yield
\beq\label{master0}
\nabla^2 \Phi = - \sqrt{\frac{\epsilon}{2}} \frac{H}{M_P} \left( v' -
\frac{z'}{z} v \right)
\eeq
where $z \equiv a \phi_0'/\mH$. Equation \eqref{master0} is expressed in terms
of gauge-invariant quantities. Nevertheless, in the longitudinal gauge, $\Phi$
represents the curvature perturbation and is related to $v$ exactly in the same
way as described in \eqref{master0}. Under the slow-roll approximation, it
leads to the following useful expressions: $z'/z\simeq
(1+2\epsilon-\delta)/(-\eta)$ and $a'/a\simeq (1+\epsilon)/(-\eta)$, similarly
$z''/z\simeq (2+6\epsilon-3\delta)/\eta^2$ and
$a''/a\simeq(2+3\epsilon)/\eta^2$.

\subsection{Quantization of the perturbations}\label{scalarquant}

As mentioned in the Introduction, the CSL model is based on a non-linear
modification to the Schr\"{o}dinger equation. Therefore, it is convenient to
begin by presenting the theory of the inflaton in the Schr\"{o}dinger picture,
where the relevant theory objects are the Hamiltonian and the wave functional.
One starts with the action of a scalar field, i.e. the inflaton, minimally
coupled to gravity, and then by expanding up to second order in the scalar
perturbations one can express the action in terms of the Mukhanov-Sasaki
variable $v(\eta,\x)$ \cite{mukhanov1992}. The resulting action, expressed in
Fourier modes
of the field $v(\eta,\x)$, reads $\delta^{(2)}S =\frac{1}{2} \int d\eta\: d^3
\nk \: \mathcal{L}$, where
\beq\label{accionv}
\mathcal{L} =  v_{\nk}' v_{\nk}^{\star '} - k^2 v_{\nk} v_{\nk}^{\star}
-\frac{z'}{z}  \left(  v_{\nk} v_{\nk}^{\star '}  +  v_{\nk}'  v_{\nk}^{\star}
\right) +  \left( \frac{z'}{z} \right)^2 v_{\nk} v_{\nk}^{\star}.
\eeq
The canonical conjugated momentum associated to $v_{\nk}$ is $p_{\nk} =
\partial
\mathcal{L}/ \partial v_{\nk}^{\star '}$, i.e.
\beq\label{momento}
 p_{\nk} =  v_{\nk}'-(z'/z)v_{\nk},
\eeq
Since $v(\eta,\x)$ is a real field, note that $v_{\nk}^\star = v_{-\nk}$.
Therefore, the Hamiltonian is $H= \frac{1}{2} \int d^3 \nk_{}\: (H_{\nk}^R +
H_{\nk}^I)$, with
\barr\label{ham}
H^{R,I}_{\nk} &=& p_{\nk}^{R,I} p_{\nk}^{\star R,I} + \frac{z'}{z} \left(
v_{\nk}^{R,I} p_{\nk}^{\star R,I} + v_{\nk}^{\star R,I} p_{\nk}^{R,I} \right)
+\nonumber \\
&+& k^2 v_{\nk}^{R,I} v_{\nk}^{\star R,I},
\earr
where the indices $R,I$ denote the real and imaginary parts of $v_{\nk}$ and
$p_{\nk}$. We now promote $v_{\nk}$ and $p_{\nk}$ to quantum operators, by
imposing canonical commutations relations $[\hat{v}_{\nk}^{R,I},
\hat{p}_{\nk'}^{R,I}] = i\delta (\nk-\nk')$.

In the Schr\"{o}dinger picture, the wave functional $\Psi[v(\eta,\x)]$
characterizes the
state of the system. Moreover, in Fourier space, the wave functional can be factorized
into mode component $\Psi[v(\eta,\nx)] = \Pi_{\nk}
\Psi_{\nk}^R (v_{\nk}^R) \times$  $\Psi_{\nk}^I (v_{\nk}^I)$. From now on, we will deal
with each mode separately.  Henceforth, each mode of the wave functional,
associated to the real and imaginary parts of the canonical variables,
satisfies
the Schr\"{o}dinger equation $\hat H_{\nk}^{R,I} \Psi_{\nk}^{R,I} = i
\partial\Psi_{\nk}^{R,I}/\partial \eta$, with the Hamiltonian provided by
\eqref{ham}. The usual assumption is that at an early time $\tau$ (i.e. the
onset of inflation), the modes are in their adiabatic ground state, which is a
Gaussian centered at zero with certain spread. Since the initial quantum state
is Gaussian, its form is preserved during the time evolution. For reasons that
will become evident in the following, it is convenient to work in the momentum
representation; thus, the Gaussian state
\beq\label{psionda}
\Psi^{R,I}(\eta,p_{\nk}^{R,I}) = \exp[-A_{k}(\eta)(p_{\nk}^{R,I})^2 +
B_{k}(\eta)p_{\nk}^{R,I} + C_{k}(\eta)]
\eeq
evolves according to Schr\"{o}dinger equation, with initial conditions given by
$A_k (\tau) = 1/2k, B_k (\tau)=C_k(\tau)=0$ corresponding to the Bunch-Davies
vacuum, which is perfectly homogeneous and isotropic in the sense of a vacuum
state in quantum field theory (see Appendix \ref{appA}).

In this work, the main reasons for invoking the self-induced collapse of the
wave function are precisely to break the homogeneity and isotropy of the initial
state, and that the emergence of the seeds of cosmic structures can be
achieved.
We will accomplish this goal by using the CSL collapse mechanism, and in the
following subsection we will provide a very brief general description of it.
For a complete review of the CSL model see for instance \cite{bassi}.

\subsection{CSL collapse mechanism}

The CSL collapse mechanism is based on a stochastic non-linear modification of
the Schr\"{o}dinger equation. Ideally, this modification induces a collapse of
the wave function toward one of the possible eigenstates of an operator $\hat
\varTheta$, called the collapse operator, with certain rate $\lambda$. The
self-induced collapse is due to the interaction of the system with a background
noise $W(t)$ that can be considered as a continuous-time stochastic process of
the Wiener kind.  The modified Schr\"{o}dinger equation drives the time
evolution of an initial state as
\beq\label{csl}
| \Psi, t \ket = \hat T \exp\{-\int_{t_0}^t dt'[i \hat H + \frac{1}{4\lambda}
(W(t')-2\lambda \hat \varTheta )^2 ]         \} |\Psi, t_0 \ket,
\eeq
with $\hat T$ the time-ordering operator. The probability associated with a
particular realization of $W(t)$ is,
\beq\label{prob}
P[W(t),t] DW(t) = \bra \Psi, t | \Psi, t \ket \prod_{t_i=t_0}^t
\frac{dW(t_i)}{\sqrt{2\pi \lambda/dt}}.
\eeq
The norm of the state $|\Psi, t\ket$ evolves dynamically, and Eq. \eqref{prob}
implies that the most probable state will be the one with the largest norm.
From \eqref{csl} and \eqref{prob}, it can be derived the evolution equation of
the density matrix operator $\hat \rho$, i.e.
\beq\label{rhodt}
\frac{d \hat \rho}{dt} = -i [\hat H, \hat \rho] - \frac{\lambda}{2} [\hat
\varTheta, [\hat \varTheta, \hat \rho]].
\eeq
The density matrix operator can be used to obtain the ensemble average of the
expectation value of an operator $\overline{\bra \hat O \ket}=\textrm{Tr}[\hat
O
\hat \rho]$. Consequently, from \eqref{rhodt} it follows that:
\beq\label{operadorcsl}
\frac{d}{dt} \overline{ \bra \hat O \ket } = -i \overline{[\hat O, \hat H]} -
\frac{\lambda}{2} \overline{[\hat \varTheta, [\hat \varTheta, \hat O]]}.
\eeq

\subsection{CSL and inflation}\label{cslinflation}

In order to apply the CSL to the inflationary regime, we need to establish which
are the appropriate observables that emerge from the quantum theory of
inflation. A reasonable observable is the curvature scalar perturbation, which
is directly related to the temperature anisotropies of the CMB, and in the
longitudinal gauge corresponds exactly to the Bardeen potential $\Phi$. A
quantization of the Mukhanov-Sasaki variable $v$ yields automatically a
quantization of $\Phi$; therefore, the question that arises here is: what is
exactly the relation between the quantum and classical objects?  In particular,
what is the relation between $\hat \Phi$ and $\Phi$?

To illustrate the point of view we will adopt, let us focus on the temperature
anisotropies of the CMB observed today on the celestial
two-sphere and its relation to the scalar metric perturbation $\Phi$. Such a relation is
approximately given by  (i.e. for large angular scales)
\begin{equation}\label{deltaT}
\frac{\delta T}{T_0} (\theta,\varphi) \simeq \frac{1}{3} \Phi.
\end{equation}
On the other hand, the observational data are  described  in terms of the
coefficients  $a_{lm}$ of the multipolar series expansion $\delta T/T_0
(\theta,\varphi)=\sum_{lm} a_{lm}Y_{lm}(\theta,\varphi)$, hence

\beq
a_{lm}=\int
\frac{\delta T}{T_0}(\theta,\varphi)Y^*_{lm}(\theta,\varphi)d\Omega,
\end{equation}
here $\theta$ and $\varphi$ are the coordinates on the celestial two-sphere,
with $Y_{lm}(\theta,\varphi)$ as the spherical harmonics.

Given Eq. \eqref{deltaT}, the  coefficients $a_{lm}$ can be further re-expressed in
terms of the Fourier modes associated to $\Phi$, i.e.

\begin{equation}\label{alm2}
a_{lm} = \frac{4 \pi i^l}{3}   \int \frac{d^3{k}}{(2 \pi)^3} j_l (kR_D)
Y_{lm}^* (\hat{k}) \Delta (k) \Phi_{\vec{k}},
\end{equation}
with $j_l (kR_D)$ the spherical Bessel function of order $l$ of the first kind and
$R_D$ is
the comoving radius of the last scattering surface. We have
explicitly included the modifications associated with late-time physics encoded
in the transfer functions $\Delta (k)$. The metric perturbation $\Phi_{\nk}$ is
the primordial curvature perturbation.

Now, how to relate $\Phi_{\nk}$ that appears in Eq. \eqref{alm2}
with the quantum operator $\hat \Phi_{\nk}$ coming from the quantum theory?
Evidently, if we compute the expectation value $\langle \hat{\Phi}_{\nk} \rangle$ in
the vacuum state $| 0  \rangle$ and identify it exactly with  $\Phi_{\nk}$, then we
obtain exactly zero; while it is clear that for
any given $l, m$, the measured value of the quantity $a_{lm}$ is not zero. As matter of
fact, the standard argument is that it is not the quantity $a_{lm}$ that is zero but the
average $\overline{a_{lm}}$. However, the notion of average is subtle, since in the CMB
one has an average over different directions in the sky, while the average that one
normally associates to the quantum expectation value of an operator is related to an
average over possible outcomes of repeatedly measurements of an observable associated to
an operator in the Hilbert space of the system (clearly the concepts of measurements,
observers, etc. are not well defined in the early universe).

Nevertheless, in the standard approach (by invoking decoherence,
squeezing of the vacuum, many-world interpretation of quantum mechanics, etc. although
we do not subscribe to such postures for the reasons exposed in Ref. \cite{shortcomings})
somehow one can make the association $\Phi_{\nk}  = A e^{i \alpha_{\nk}}$ with
$\alpha_{\nk}$ a random
phase and $A$ being identified with the quantum uncertainty of $\hat \Phi_{\nk}$,
i.e. $A^2=\bra 0 | \hat \Phi_{\nk}^2 |0\ket$. But the random nature of
$\Phi_{\nk}$, codified in the random number $e^{i \alpha_{\nk}}$, remains unclear.
In fact, the relation $\Phi_{\nk}  = A e^{i \alpha_{\nk}}$ is not valid at all times in
the standard approach. It is only valid after the proper wavelength of the mode is larger
than the Hubble radius (or when  ``the mode has crossed the horizon''). That is,
the traditional statement is that during inflation the modes become super-horizon and
then occurs the transition $\hat \Phi \to \Phi$. Furthermore, the relation $\Phi_{\nk}
 =  \sqrt{\bra 0 | \hat \Phi_{\nk}^2 |0\ket}e^{i \alpha_{\nk}}$ could generate some
misunderstandings, in the sense that one might think that, since the vacuum fluctuations
$\bra 0 | \hat \Phi_{\nk}^2 |0\ket$ are not zero, $\Phi_{\nk}$ should be different
from zero when the mode is super-horizon, and the spacetime perturbations are ``born''
(see Appendix \ref{appA} for clarification).

On the other hand, in our approach, the random nature of $\Phi_{\nk}$ will
come directly
from the stochastic aspects of the quantum dynamical reduction, i.e. from the
CSL mechanism. Moreover, we will adopt the point of view that the classical
characterization of  $\Phi$
is an adequate description if the quantum state is sharply peaked around some
particular value. In consequence, the classical value corresponds to the
expectation value of $\hat \Phi$ \cite{gabriel}. More precisely, the CSL
collapse mechanism will lead to a final state such that the relation
\beq\label{igualdadchingona}
\Phi_{\nk} =  \bra \Psi| \hat \Phi_{\nk} |  \Psi \ket
\eeq
is valid.

Therefore, in our approach, the coefficients $a_{lm}$  in Eq. \eqref{alm2}, will be
given by

\begin{equation}\label{alm3}
a_{lm} = \frac{4 \pi i^l}{3}   \int \frac{d^3{k}}{(2 \pi)^3} j_l (kR_D)
Y_{lm}^* (\hat{k}) \Delta (k) \bra \Theta | \hat{\Phi}_{\vec{k}} | \Theta \ket,
\end{equation}
where $| \Theta \ket$ corresponds to the evolved state according to the
non-unitary modification of the Schr\"{o}dinger equation provided by the CSL
mechanism. At this point, we encourage the reader to consult Appendix \ref{appB} for a
discussion of other possible ways to relate $\hat \Phi$ and $\Phi$ within the CSL
framework.

Additionally, Eqs. \eqref{master0} and \eqref{igualdadchingona}
allow us to relate each mode of the curvature perturbation to the quantum field
variables. That is,
\beq\label{igualdadchingona2}
\Phi_{\nk} = \bra \hat \Phi_{\nk} \ket =  \sqrt{\frac{\epsilon}{2}}
\frac{H}{k^2
M_P} \bra \hat p_{\nk} \ket = \sqrt{\frac{\epsilon}{2}} \frac{H}{k^2 M_P} [\bra
\hat p_{\nk}^R \ket+i \bra \hat p_{\nk}^I \ket],
\eeq
where we used the definition of the momentum provided by \eqref{momento}.
Equation \eqref{igualdadchingona2} relates the classical curvature perturbation
with the momentum variable of the quantum field, consequently, this strongly
suggest that the collapse operator $\hat \varTheta$ to be considered in the CSL
mechanism is the momentum operator. Thus, if the initial state is the
Bunch-Davies vacuum state $\bra p_{\nk}^{R,I}| 0, \tau \ket \propto
\exp[-(p_{\nk}^{R,I})^2/2k]$, then $\bra0| \hat p_{\nk}^{R,I} |0\ket = 0$, and
as a consequence of \eqref{igualdadchingona2}, the curvature perturbation is
$\Phi_{\nk}=0$, i.e. the spacetime is perfectly homogeneous and isotropic. It
is only after the state has evolved, according to the CSL mechanism, that
generically $\bra  \hat p_{\nk}^{R,I} \ket \neq 0$ and the curvature
perturbation is born. This illustrates how the self-induced collapse provided
by the CSL model can generate the primordial perturbations.

Furthermore, from the previous discussion, it is evident that we will adopt the
CSL mechanism for each mode of the field and the corresponding real and
imaginary parts. Therefore, the evolution of the state vector characterizing
each mode of the inflaton field, written in conformal time, will be
\barr\label{cslmodos}
| \Psi_{\nk}^{R,I}, \eta \ket &=& \hat T \exp\{-\int_{\tau}^\eta d\eta'[i \hat
H_{\nk}^{R,I} +\nonumber \\
&+& \frac{1}{4\lambda_{k}} (W(\eta')-2\lambda_{k} \hat p_{\nk}^{R,I} )^2 ]  \}
|\Psi_{\nk}^{R,I}, \tau \ket
\earr
with $H_{\nk}^{R,I}$ given in \eqref{ham}. Moreover, with the previous
considerations, the motion equation \eqref{operadorcsl} for the ensemble
average
of the expectation value of an operator is:
\beq\label{operadorcslmodos}
\frac{d}{d\eta} \overline{ \bra \hat O_{\nk}^{R,I} \ket } = -i \overline{[\hat
O_{\nk}^{R,I}, \hat H_{\nk}^{R,I}]} - \frac{\lambda_k}{2} \overline{[\hat
p_{\nk}^{R,I}, [\hat p_{\nk}^{R,I}, \hat O_{\nk}^{R,I}]]}.
\eeq
This completes our treatment of the CSL model during inflation.

\subsection{The scalar power spectrum}

Having established the relation between the objects $\hat \Phi$ and $\Phi$, we
now focus on the scalar power spectrum. The scalar power spectrum in Fourier
space is defined as
\beq\label{psdef}
\overline{\Phi_{\nk} \Phi_{\nk'}^\star} \equiv \frac{2\pi^2}{k^3} \mP_s (k)
\delta(\nk-\nk')
\eeq
where $\mP_s (k)$ is the dimensionless power spectrum. The bar appearing in
\eqref{psdef} denotes an ensemble average over possible realizations of the
stochastic field $\Phi_{\nk}$. In our approach, the realization of a particular
$\Phi_{\nk}$ is given by the self-induced collapse that results from the CSL
mechanism.

Henceforth, Eq. \eqref{igualdadchingona2} implies that $\overline{\Phi_{\nk}
\Phi_{\nk'}^\star} \propto$
$\overline{\bra \hat{p}_{\nk}  \ket \bra
\hat{p}_{\nk'} \ket^\star }$, where the expectation values are being evaluated
at the (evolved) state provided by \eqref{cslmodos}. More explicitly,
\barr\label{2puntosmomento}
\overline{\bra \hat{p}_{\nk}  \ket \bra \hat{p}_{\nk'} \ket^\star } &=&
\overline{\bra \hat p_{\nk}^R+i \hat p_{\nk}^I\ket \bra \hat p_{\nk'}^R-i \hat
p_{\nk'}^I\ket} \nonumber \\
&=& \left( \overline{\bra \hat p_{\nk}^R \ket^2} + \overline{ \bra \hat
p_{\nk}^I \ket^2 } \right) \delta(\nk-\nk'),
\earr
where in the second line we assumed that the CSL model does not induce modes
correlations. Thus, from now on we will focus on calculate the quantities
$\overline{\bra \hat p_{\nk}^{R,I} \ket^2}$. In particular, we will calculate
only the real part since the computation for the imaginary part proceeds in the
same fashion. In order to simplify the notation, we will omit the index $R$
unless it can create confusion, in which case we will write it explicitly.

By using the Gaussian wave function in the momentum representation
\eqref{psionda}, and the probability associated to $W(\eta)$ \eqref{prob}, it
can be shown that \cite{pedro},
\beq\label{resta}
\overline{\bra \hat p_{\nk} \ket^2} = \overline{\bra \hat p_{\nk}^2 \ket} -
\frac{1}{4 \text{Re}[A_k (\eta)]}.
\eeq
That is, $(4 \text{Re}[A(\eta)])^{-1}$ is the standard deviation of the squared
momentum. It is also the width of every packet in momentum space. Thus, to
calculate $\overline{\bra \hat p_{\nk} \ket^2}$, we only need to find the two
terms on the right hand side of \eqref{resta}. The second term will be found
from the CSL evolution Eq. \eqref{cslmodos}, and the first one by using
Eq. \eqref{operadorcslmodos}.

Let us focus on the second term. Using the general Gaussian state in momentum
space \eqref{psionda} and the CSL evolution Eq. \eqref{cslmodos} for the
wave function, the equation of motion for $A_k (\eta)$ results
\beq
A_k' = \frac{i}{2} + \lambda_k  + 2 A_k \frac{z'}{z} - 2ik^2 A_k^2.
\eeq
The previous equation can be solved by performing the change of variable
$A_k(\eta) \equiv f'(\eta)/[2ik^2 f(\eta)]$, resulting in a Bessel differential
equation for $f$. After solving such an equation, and returning to the original
variable $A_k$, we obtain
\beq\label{Aexacta}
A_k(\eta) = \frac{q}{2ik^2} \left[\frac{J_{\nu_s+1} (-q\eta)+e^{-i\pi\nu_s}
J_{-(\nu_s+1)} (-q\eta)}{J_{\nu_s} (-q\eta) -e^{-i\pi\nu_s} J_{-\nu_s}
(-q\eta)}
      \right],
\eeq
being $q^2 \equiv k^2 (1-2i\lambda_k)$ and where the initial condition for the
Bunch-Davis vacuum $A_k(\tau) = 1/2k$ was used (recall that $\tau$ corresponds
to the onset of inflation, thus, $\tau \to -\infty$). $J_{\nu_s}$ corresponds to a
Bessel function of the first kind of order $\nu_s = 1/2 + 2 \epsilon - \delta$.
Note that Eq. \eqref{Aexacta} is exact. However, from the observational
point of view, one is interested in modes that are well outside the Hubble
radius during inflation, i.e. modes with $k \ll aH$ or equivalently, $-k\eta
\to
0$. Therefore, we can expand $A_k(\eta)$ in the limit $-q\eta \to 0$ (provided
that $\lambda_k \ll 1$):
\beq\label{Aaprox}
A_k(\eta) \simeq \frac{i \nu_s}{k^2\eta} \bigg[ 1 + (-q\eta)^{2\nu_s} \nu_s
2^{-2\nu_s} \frac{\Gamma(-\nu_s)}{\Gamma(\nu_s+1)}  e^{i \pi \nu_s} \bigg]^{-1},
\eeq
with $\Gamma(\nu)$ the Gamma function. From \eqref{Aaprox}, it is
straightforward to obtain
\barr\label{ReA}
& & \frac{1}{4\textrm{Re}[A_k(\eta)]} \simeq \nn
 & & \frac{k 2^{2\nu_s-2} \zeta_k^{-2\nu_s}
\sin (\pi \nu_s) \Gamma^2 (\nu_s) (-k\eta)^{-2\nu_s+1} }{\pi \sin (2\nu_s
\theta_k + \pi \nu_s)}
\earr
where we have defined $\zeta_k e^{i\theta_k} \equiv \sqrt{1-2i\lambda_k}$.

Next, we focus on the first term of \eqref{resta}, i.e.  $\overline{\bra \hat
p_{\nk}^2 \ket}$. It will be useful to define the quantities $Q
\equiv
\overline{\bra \hat v_{\nk}^2 \ket}$, $R\equiv \overline{\bra \hat p_{\nk}^2
\ket}$ and $S\equiv \overline{\bra \hat p_{\nk} \hat v_{\nk} + \hat v_{\nk}
\hat
p_{\nk} \ket}$. Thus, the evolution equations for $Q, R$ and $S$ are obtained
using \eqref{operadorcslmodos}:
\beq\label{QRS}
Q' = S+2\frac{z'}{z}+\lambda_k, \enskip R'=-k^2 S -2 \frac{z'}{z}R, \enskip
S'=2R - 2k^2 Q.
\eeq
Therefore, we have a linear system of coupled differential equations, whose
general solution is a particular solution to the system plus a solution to the
homogeneous equation (with $\lambda_k=0$). After a long series of calculations
we find
\barr\label{R}
R (\eta) &=& k^2 (-k \eta) \bigg[C_1 J^2_{\nu_s}(-k\eta) + C_2 J_{-\nu_s}^2
(-k\eta) +\nonumber \\
&+& C_3 J_{\nu_s} (-k\eta) J_{-\nu_s} (-k\eta)  \bigg] + \frac{\lambda_k k^2
\eta}{4 \nu_s}
\earr
where the constants $C_1, C_2$ and $C_3$ are found by imposing the initial
conditions corresponding to the Bunch-Davies vacuum state: $Q(\tau)=1/2k,
R(\tau) = k/2$, $S(\tau)=0$. Equation \eqref{R} is exact; expanding it again
around $-k\eta \to0$ yields
\barr\label{Raprox}
R(\eta) &\simeq&  \frac{k}{\pi} 2^{2\nu_s-2} \Gamma^2 (\nu_s) \bigg[1+
\lambda_k
\sin \beta_k \cos \beta_k -  \frac{\lambda_k k \tau}{2} \nonumber \\
 &\times& \left(\frac{3}{\nu_s+1} \sin^2 \beta_k + \frac{\cos^2 \beta_k}{\nu_s}
\right)  \bigg]  (-k\eta)^{-2\nu_s+1}, \nonumber \\
\earr
with $\beta_k \equiv -k\tau -\nu_s \pi/2-3\pi/4$. Also, recall that $R(\eta)
\equiv  \overline{\bra \hat p_{\nk}^2 \ket}$ .

At this point, we have the two terms of \eqref{resta}, i.e. $ \overline{\bra
\hat p_{\nk}^2 \ket}$ from Eq. \eqref{Raprox}, and
$1/(4\textrm{Re}[A_k(\eta)])$ from Eq. \eqref{ReA}. Then we can write
\beq\label{p2}
\overline{\bra \hat p_{\nk} \ket^2} \simeq \frac{k}{\pi} 2^{2\nu_-2}
\Gamma^2(\nu_s) (-k\eta)^{-2\nu_s+1} F(\lambda_k,\nu_s),
\eeq
where we have defined
\barr\label{F}
F(\lambda_k, \nu_s) &\equiv& \bigg[1+ \lambda_k \sin \beta_k \cos \beta_k  -
\frac{\lambda_k k \tau}{2} \bigg(\frac{3}{\nu_s+1}  \nonumber \\
 &\times& \sin^2 \beta_k +  \frac{\cos^2 \beta_k}{\nu_s} \bigg)-
\frac{\zeta_k^{-2\nu_s} \sin (\nu_s \pi)}{\sin(2\nu_s \theta_k + \nu_s \pi)}
\bigg]. \nonumber \\
\earr
Equation \eqref{p2} is valid for both the real and the imaginary part of $\hat
p_{\nk}$. Therefore, substituting \eqref{p2} into \eqref{2puntosmomento}, and
taking into account \eqref{igualdadchingona2}, we obtain
\barr\label{2ptosphi}
\overline{\Phi_{\nk} \Phi_{\nk'}^\star} &=& \frac{\epsilon H^2}{k^3 \pi M_P^2}
2^{2\nu_-2} \Gamma^2(\nu_s) (-k\eta)^{-2\nu_s+1} F(\lambda_k,\nu_s)   \nonumber
\\
&\times& \delta(\nk-\nk')
\earr

Finally, with the definition of the power spectrum \eqref{psdef}, and the
result
in \eqref{2ptosphi}, the scalar power spectrum within the CSL model,
results\footnote{Note that in Eq. \eqref{2ptosphi} the slow-roll parameter
$\epsilon$ appears in the numerator, while in the expression for the scalar
power spectrum \eqref{scalarps} appears in the denominator. The reason for
this difference is that, in the longitudinal gauge, the scalar curvature
perturbation $\Phi$ becomes amplified by a factor of $1/\epsilon$ during the
transition from inflation to the radiation dominated stage
\cite{gabriel2010,deruelle}, in which the
CMB is created (the decoupling era). Therefore in order to obtain a consistent
prediction to be compared with the observational data, we must multiply by a
factor of $1/\epsilon^2$ the scalar power spectrum obtained during inflation
associated to $\overline{\Phi_{\nk} \Phi_{\nk'}^\star}$.}
\beq\label{scalarps}
\mP_s (k) = \frac{H^2 2^{2\nu_s-3} \Gamma^2 (\nu_s)}{\epsilon M_P^2 \pi^3}
(-k\eta)^{-2\nu_s+1} F(\lambda_k,\nu_s)
\eeq

We will discuss some physical implications from our prediction in Sect.
\ref{discusion}.

\section{The CSL model and the tensor power spectrum}
\label{Sectensor}
Once we have successfully applied the CSL model to the primordial scalar
perturbations, we now proceed to focus on the tensor perturbations. As it is well known,
these
perturbations represent gravitational waves characterized by a traceless,
transverse and symmetric tensor field. These properties imply that the
gravitational waves are polarized in two ways. As is traditional, we will
consider one type of polarization and at the end we will just multiply the final results by a
factor of 2.

The action for the tensor perturbations is obtained from the Einstein-Hilbert
action by expanding the tensor perturbations $h_{ij} (\x,\eta)$ up to
second order \cite{mukhanov1992}. The resulting action for the tensor field
$h_{ij}
(\x,\eta)$ can be expressed in terms of its Fourier modes $h_{ij}
(\nk,\eta)=h_{\nk} (\eta) e_{ij} (\nk)$, with $e_{ij} (\nk)$ representing a
time-independent polarization tensor. Performing the change of variable
\begin{equation}\label{cambiovariable}
h_{\nk} (\eta) \equiv \frac{2}{M_P (e^i_{\,j}e^j_{\,i})^{1/2}} \frac{v_{\nk}
(\eta)}{a(\eta)},
\end{equation}
the action can be written as $\delta^{(2)} S_h =\frac{1}{2} \int d\eta \:d^3
\nk \:\mathcal{L}_h$, where
\beq\label{accionh}
\mathcal{L}_h =  v_{\nk}' v_{\nk}^{\star '} - k^2 v_{\nk} v_{\nk}^{\star}
-\frac{a'}{a}  \left(  v_{\nk} v_{\nk}^{\star '}  +  v_{\nk}'  v_{\nk}^{\star}
\right) +  \left( \frac{a'}{a} \right)^2 v_{\nk} v_{\nk}^{\star}.
\eeq

By comparing the action in \eqref{accionv} with the one obtained for the tensor
perturbations \eqref{accionh}, we see that they are equivalent as expected. The
only difference is that the action for the scalar perturbations contains a term
$z'/z$ while the action for the tensor modes contains the term $a'/a$.
Therefore, the quantization procedure of Sect. \ref{scalarquant} remains
exactly the same by simply replacing $z'/z \to a'/a$. Now, as was done in
Sect. \ref{cslinflation}, we need to define which are the appropriate
observables that emerge from the quantum theory of the tensor modes. By the
same
argument that allowed us to identify $\Phi_{\nk} = \bra \hat \Phi_{\nk} \ket$,
we will assume that
\beq\label{igualdadchingonah}
h_{\nk} = \bra \Psi | \hat h_{\nk} | \Psi \ket
\eeq
is also valid. Hence, it is evident that a quantization of $v_{\nk}$ from the
action \eqref{accionh} yields a quantization of $\hat h_{\nk}$. In other words,
Eqs. \eqref{cambiovariable} and \eqref{igualdadchingonah} imply that
\beq\label{igualdadchingona2h}
h_{\nk} (\eta) = \frac{2}{M_P (e^i_{\,j}e^j_{\,i})^{1/2}} \frac{\bra \hat
v_{\nk} (\eta) \ket}{a(\eta)}.
\eeq
Thus, \eqref{igualdadchingona2h} relates the quantum field variable $\hat
v_{\nk}$ to the amplitude of the tensor mode $h_{\nk}$. Furthermore, this
relation is analogous to \eqref{igualdadchingona2}, in which the scalar
curvature perturbation is related to the momentum field variable $\hat p_{\nk}$.

On the other hand, as a matter of physical consistency in the operation of the
CSL model, we will keep the assumption that the collapse operator corresponds
to
the momentum operator, even if the relation between $h_{\nk}$ and the quantum
matter fields is in terms of $v_{\nk}$. In other words, the quantum theory
described by actions \eqref{accionh} and \eqref{accionv} are physically
equivalent, i.e. they correspond to the action of a scalar field with time
dependent mass. Thus, the fields 'do not know' how they will be related to the
curvature perturbation. We will deepen this discussion in the next section.

Therefore, since we retained the momentum operator as the collapse operator,
the treatment of the CSL model, applied to inflation, remains the same as the one
presented in \eqref{cslmodos} and \eqref{operadorcslmodos}. Moreover, given
that the quantization procedure of Sect. \ref{scalarquant} stayed unchanged, we
will continue working with the wave function shown in \eqref{psionda}.

With the relation between $\hat h_{\nk}$ and $h_{\nk}$ already established, we
can now focus on the dimensionless tensor power spectrum defined as
\beq\label{pshdef}
\overline{h^i_j(\nk) h^{j\star}_i (\nk')}  \equiv \frac{2\pi^2}{k^3} \mP_t (k)
\delta (\nk-\nk').
\eeq
Furthermore, by using \eqref{igualdadchingona2h} it results in
\beq\label{ensambleh}
\overline{h^i_j(\nk) h^{j\star}_i (\nk')} \propto \overline{\bra \hat{v}_{\nk}
\ket \bra \hat{v}_{\nk'} \ket^\star }=\left( \overline{\bra \hat v_{\nk}^R
\ket^2} + \overline{ \bra \hat v_{\nk}^I \ket^2 } \right) \delta(\nk-\nk'),
\eeq
Proceeding in a similar manner with which we arrive at \eqref{resta}, we now
find that
\beq\label{restah}
\overline{\bra \hat v_{\nk}^{R,I} \ket^2} = \overline{\bra (\hat
v_{\nk}^{R,I})^2 \ket} - \frac{|A_k(\eta)|^2}{ \text{Re}[A_k (\eta)]}.
\eeq
The second term of this expression is obtained from Eq. \eqref{Aaprox}, that
is,
in the limit $-k\eta \to 0$ we can write
\beq\label{nose}
\frac{|A_k(\eta)|^2}{ \text{Re}[A_k (\eta)]} \simeq \frac{\sin \pi \nu_t
\Gamma^2(\nu_t+1)2^{2\nu_t} \zeta_k^{-2\nu_t} (-k\eta)^{-2\nu_t-1}}{k
\sin(2\nu_t \theta_k + \pi \nu_t) \pi},
\eeq
where $\nu_t = 1/2 + \epsilon$. This is different from
$\nu_s=1/2+2\epsilon-\delta$. This distinction arises because we have replaced
$z'/z \to a'/a$ in the whole computation.

Next, in order to obtain the first term in \eqref{restah}, i.e. $\overline{\bra (\hat
v_{\nk}^{R,I})^2 \ket}$, we note that its evolution equation is already shown
in
the system of differential equations given by \eqref{QRS}, recalling that $Q
\equiv  \overline{\bra (\hat v_{\nk}^{R,I})^2 \ket}$ (and also that $z'/z \to
a'/a$). The solution for $Q$ can be found analytically. However, since we are
interested in the modes such that $-k\eta \to 0$, $Q(\eta)$ becomes
\barr\label{Qaprox}
Q(\eta) &\simeq&  \frac{2^{2\nu_t}}{k \pi}  \Gamma^2 (\nu_t+1) \bigg[1+
\lambda_k \sin \beta_k \cos \beta_k -  \frac{\lambda_k k \tau}{2} \nonumber \\
 &\times& \left(\frac{3}{\nu_t+1} \sin^2 \beta_k + \frac{\cos^2 \beta_k}{\nu_t}
\right)  \bigg]  (-k\eta)^{-2\nu_t-1}, \nonumber \\
\earr
with $\beta_k$ being the same as in \eqref{Raprox} but replacing $\nu_s \to
\nu_t$. By using \eqref{nose} and \eqref{Qaprox} into \eqref{restah} we write
\beq\label{ensamblev}
\overline{\bra \hat v_{\nk}^{R,I} \ket^2} \simeq \frac{2^{2\nu_t}}{k \pi}
\Gamma^2 (\nu_t+1) (-k\eta)^{-2\nu_t-1} F(\lambda_k,\nu_t).
\eeq
Consequently, Eq. \eqref{ensamblev} along with \eqref{ensambleh} imply that
\barr\label{ensambleh2}
\overline{h^i_j(\nk) h^{j\star}_i (\nk')} &=&  \frac{8}{M_p^2 a^2}
\frac{2^{2\nu_t}}{k \pi}  \Gamma^2 (\nu_t+1) (-k\eta)^{-2\nu_t-1} \nonumber \\
&\times& F(\lambda_k,\nu_t) \delta(\nk-\nk').
\earr
Now, by comparing \eqref{ensambleh2} with the definition of the tensor power
spectrum \eqref{pshdef}, we finally obtain
\barr\label{tensorps}
\mP_t (\eta) &=& \frac{ H^2     }{ M_P^2 \pi^3     } (-\eta)^{-2\nu_t+1}
2^{2\nu_t+3} \Gamma^2 (1+\nu_t) \nonumber \\
&\times&k^{-2\nu_t+1}  F(\lambda_k,\nu_t),
\earr
where we have multiplied by a factor of 2 due to the polarization of the
gravitational waves, and we used $a(\eta) \simeq -1/(H\eta)$. With the
expressions of the scalar and tensor power spectrum at hand, Eqs.
\eqref{scalarps} and \eqref{tensorps} respectively, it is now straightforward
to calculate the tensor-to-scalar ratio defined as $r\equiv \mP_t (k)/\mP_s (k)$.
Under the approximation $\nu_s \simeq \nu_t \simeq 1/2$, it leads to
$r\simeq 16 \epsilon$, which is exactly the same prediction as in the traditional
inflationary scenario. This result is also consistent with the one obtained in
\cite{mauro} in which the tensor-to-scalar ratio remained unchanged when
considering a generic collapse scheme.

\section{Discussion}
\label{discusion}

With the power spectra within the CSL framework calculated, we will make a few
remarks. Let us rewrite the scalar and tensor power spectra in the following
suggestive forms:
\beq\label{psfinal}
\mP_s(k) = A_s k^{n_s-1} F(\lambda_k,\nu_s) \:,\:\: \mP_t(k) = A_t k^{n_t}
F(\lambda_k,\nu_t)
\eeq
where we define
\begin{subequations}\label{parametros}
\beq
A_s \equiv \frac{H^2(-\eta)^{-2\nu_s+1} 2^{2\nu_s-3} \Gamma^2 (\nu_s)}{\epsilon
M_P^2 \pi^3}, \enskip n_s-1 \equiv -2\nu_s+1,
\eeq
\beq
A_t \equiv \frac{ H^2   (-\eta)^{-2\nu_t+1} 2^{2\nu_t+3} \Gamma^2 (\nu_t+1) }{
M_P^2 \pi^3     }, \enskip n_t \equiv -2\nu_t+1.
\eeq
\end{subequations}

The quantities, $A_s$ and $A_t$ correspond to the amplitude of the scalar and
tensor power spectrum respectively. From the definition of $A_s$, it can be
shown \cite{kinney} that this amplitude is practically a time-independent
quantity
[i.e. $d/d\eta \{ A_s\} = \mathcal{O} (\epsilon^2, \delta^2)$]. The same
argument applies for $A_t$. Actually, the amplitudes $A_s$ and $A_t$ coincide
exactly with the ones from standard predictions. Moreover, since the amplitudes
are constants, it is customary to evaluate the power spectra at some conformal
time $\eta_*$, which is usually taken to be the conformal time at the horizon
crossing $-k\eta_* = 1$. However, from our point of view, the self-induced
collapse of the wave function generates the primordial perturbations.
Therefore, we cannot evaluate the power spectra at some conformal time, e.g. at the
horizon crossing, if the collapse has not taken place yet (or more precisely the CSL
mechanism has not concluded), but this evaluation is purely conventional, our
prediction for the power spectra is time independent.

The shapes of the power spectra (that is, its dependence on $k$) are of the form
$k^{n_s-1} F(\lambda_k,\nu_s)$ and $k^{n_t} F(\lambda_k,\nu_t)$ for the scalar
and tensor power spectrum, respectively. Let us focus on the function
$F(\lambda_k,\nu)$ shown in \eqref{F} (where $\nu=\nu_s, \nu_t$), and make the
approximation $\nu_s \simeq \nu_t \simeq 1/2$, i.e.
\barr
F(\lambda_k,1/2) &=& 1+ \lambda_k [ \cos(-k\tau) \sin (-k\tau) -  k \tau]
\nonumber \\
&-& \sqrt{\frac{2}{1+\sqrt{1+4\lambda_k^2}}},
\earr
where we used the definitions of $\beta_k$, $\zeta_k$ and $\theta_k$.
Furthermore, if the following assumptions are valid: i) $-k\tau \gg 1$ and ii)
$\lambda_k \ll 1$, then
\beq
F(\lambda_k,1/2) \simeq  -\lambda_k k \tau.
\eeq
Henceforth, if $\lambda_k = \lambda_0/k$, then $F$ becomes essentially
scale-invariant and the only dependence on $k$ in the power spectra is of the
form $k^{n_s-1}$ and $k^{n_t}$.
In fact, from \eqref{parametros} and the definitions of $\nu_s$ and $\nu_t$, we
have $n_s-1 = -4\epsilon + 2 \delta$ and $n_t = -2 \epsilon$. That is, the
spectral indices are exactly the same as in the standard approach. Thus, if
$\lambda_k = \lambda_0/k$ we recover the standard prediction, what was also
noted in \cite{pedro} but within the semiclassical gravity and just for the scalar case.
In Sect. \ref{cslpar} we will say more about the
validity of i) and ii).

Finally, it is interesting to observe that if $\lambda_k=0$, i.e. the vacuum
state is evolving according to the unmodified Schr\"{o}dinger equation and no
self-induced collapse occurs, then $F(0,\nu) = 0$ (with $\nu=\nu_s,\nu_t$).
Consequently, $\mP_s=\mP_t=0$. This implies that there are no primordial
perturbations and the spacetime is perfectly homogeneous and isotropic. On the
other hand, the primordial perturbations are generated by ``switching on'' the
self-induced collapse $\lambda_k \neq 0$. In this case, $\Phi_{\nk}$ is created
randomly, and its power spectrum is obtained from $\overline{\bra
\hat{\Phi}_{\nk} \ket \bra \hat{\Phi}_{\nk'}^* \ket }$. Note that one could
argue that maybe the cancellation in the case $\lambda_k=0$ occurs only in the
limit $-k\eta\to0$, i.e. only affects the super-horizon modes. We have checked
that even if one considers the exact analytic expressions of $\overline{\bra
\hat p_{\nk}^2 \ket}$ and $1/(4 \text{Re}[A_k (\eta)])$, as well as
$\overline{\bra (\hat v_{\nk}^{R,I})^2 \ket}$ and $|A_k(\eta)|^2/(\text{Re}[A_k
(\eta)])$ in the tensor case, one arrives at the same result, that is, the
spacetime does not contain perturbations of any scale.

\section{CSL, squeezing, gauge invariance and previous works}
\label{preworks}

\subsection{CSL parameter}
\label{cslpar}
We begin this section by giving a rough estimate for the value of the CSL
parameter $\lambda_k$. As mentioned in the previous section, if assumptions i)
and ii) are valid, then by taking $\lambda_k = \lambda_0/k$ the standard
prediction for the power spectra is recovered. Assumption i) reads $-k\tau \gg
1$. An estimate for $\tau$ can be obtained by assuming that the energy scale at
the onset of inflation is $10^{16}$ GeV, and that inflation lasts $\sim 70$
e-folds. Hence, $\tau \simeq -10^{7}$ Mpc. Since the observational range for
$k$
is $10^{-6}$ Mpc$^{-1} \leq k \leq 10^{-1}$ Mpc$^{-1}$, assumption i) is
verified. On the other hand, assumption ii) is $\lambda_k = \lambda_0/k \ll 1$.
Thus, by using again the range for $k$, the CSL parameter must satisfy
$\lambda_0 \ll 10^{-6}$ Mpc$^{-1}$. We can check that this is indeed the case
form \eqref{psfinal} along with Friedmann equation. This yields $P_s(k) \propto
\frac{V}{M_P^4 \epsilon} (-\lambda_0 \tau)$ and as a consequence, if $\lambda_0
\simeq -1/\tau$, our model prediction is consistent with the standard accepted
result. Moreover, since $\tau \simeq -10^{7}$ Mpc, $\lambda_0 \simeq
10^{-7}$ Mpc$^{-1} \simeq  10^{-21}$ s$^{-1}$, in which case, assumption ii) is
valid.

Thus, if the CSL parameter is of the form $\lambda_k = \lambda_0/k$, with
$\lambda_0 \simeq  10^{-21}$ s$^{-1}$, the model is compatible with the CMB
observational data. This result is also consistent with findings of the CSL
collapse mechanism applied to situations different from the cosmological
context. In particular, the fact that $\lambda_k$ depends on the wave number
$k$ captures the spirit of the 'amplification mechanism' of the CSL model, which
states that micro- and macro-objects do not behave the same way. For example,
micro-objects tend to exhibit quantum features (e.g. superposition in their
wave functions) while macro-objects do not. In our model, the particular dependence
on $k$ of the CSL parameter (which generically sets the strength of the
self-induced collapse) implies that the self-induced collapse affects the
large-scale modes more than the low-scale modes; in some sense, the low-scale
modes behave quantum mechanically while large-scale modes can be treated
classically. It is also interesting to compare our value $\lambda_0 \simeq
10^{-21}$ s$^{-1}$ with the one suggested by the GRW model
($\lambda_{\textrm{GRW}} = 10^{-16}$ s$^{-1}$), adopted later in the CSL model
with the mass-density operator as the collapse operator \cite{pearle,pearle2}.
Our estimated value for $\lambda_0$ makes the strength of the self-induced collapse five
orders of magnitude weaker than the one in the GRW model. However, our estimate
was based on very robust assumptions regarding the inflationary era,
particularly, the energy scale of inflation and the number of e-foldings. Note
that the value of the parameter $\epsilon$ can also change the value of
$\lambda_0$ by a couple of orders of magnitude. Thus, a more careful analysis
including the full angular power spectrum of the CMB data is required in order
to constrain the value of the CSL parameter. This subject is left for a future
work.

\subsection{Squeezing of the modes}

It is usually argued that during inflation, the Bunch-Davies vacuum evolves
toward a squeezed state; that is, a state in which its uncertainty is
increased in one variable and decreased in another one, such that the product of the
uncertainties satisfy the minimum value allowed by the Heisenberg uncertainty
principle. In the Schr\"{o}dinger picture, the uncertainties of the real and
imaginary parts of $\hat p_{\nk}$ and $\hat v_{\nk}$ (associated to scalar
perturbations) are given by
\beq
 \Delta^2 \hat{p}_{\nk}^{R,I} (\eta) = \frac{1}{4 \textrm{Re} [A_k(\eta)]},
\:\:\:\:\:\:\: \Delta^2 \hat{v}_{\nk}^{R,I} (\eta) = \frac{|A_k(\eta)|^2}{
\textrm{Re} [A_k(\eta)]}.
\eeq
Henceforth, by using \eqref{ReA} and \eqref{nose}, we see that
\barr
\Delta^2 \hat{v}_{\nk}^{R,I} (\eta) &\sim& (-k\eta)^{-2\nu_s-1} \nonumber \\
\Delta^2 \hat{p}_{\nk}^{R,I} (\eta) &\sim& (-k\eta)^{-2\nu_s+1}
\earr
as $-k\eta \to 0$. Thus,
the uncertainty in $\hat{v}_{\nk}^{R,I}$ increases while the uncertainty in
$\hat{p}_{\nk}^{R,I}$ decreases as inflation takes place. This is an expected
result; in fact, it is normally used to show that inflation induces squeezing
in
the momentum variable \cite{kiefer,grishchuk,jmartin,hinduesS}. Furthermore,
recall that in our approach, the
self-induced collapse generates the primordial curvature perturbation. In the
longitudinal gauge, our point of view was reflected in Eq.
\eqref{igualdadchingona2}; therefore, the uncertainty in the momentum and in
the
curvature perturbation operator $\Phi$ are directly related, and since during
inflation $ \Delta^2 \hat{p}_{\nk}^{R,I} (\eta)$ decreases, $ \Delta^2
\hat{\Phi}_{\nk}^{R,I} (\eta)$ also decreases. On the other hand, if we had
chosen the comoving gauge, the curvature perturbation would be given by
$\mathcal{R} = v/z$, where $z \simeq a$. Consequently, in our approach, there
would be a relation between the uncertainties $\Delta^2
\hat{\mathcal{R}}_{\nk}^{R,I} (\eta) = \Delta^2 \hat{v}_{\nk}^{R,I} (\eta)/a^2
$, and since during inflation $a(\eta) \simeq -1/\eta$ then $\Delta^2
\hat{\mathcal{R}}_{\nk}^{R,I} (\eta) \sim (-k\eta)^{-2\nu+1}$. Thus, it
decreases in the same fashion as $ \Delta^2 \hat{p}_{\nk}^{R,I} (\eta)$. The
tensor modes follow this behavior, i.e. since $\hat h_{\nk} \simeq
\hat{v}_{\nk}/a$, the uncertainty in $\hat{h}_{\nk}^{R,I}$ decreases even
if the uncertainty in $\hat{v}_{\nk}^{R,I}$ increases. One could argue that
this
behavior occurs irrespectively of whether there is a quantum collapse or not. However,
it should be borne in mind that, in our approach, if there is no quantum
collapse then the classical curvature perturbation is exactly zero in spite of
its quantum uncertainty $\Delta^2 \hat{\Phi}_{\nk}^{R,I} (\eta)$ (in the
longitudinal gauge) or $\Delta^2 \hat{\mathcal{R}}_{\nk}^{R,I} (\eta)$ (in the
comoving gauge) decreases as inflation takes place. We should also mention that
even if the Bardeen potential $\Phi$, the Mukhanov-Sasaki variable $v$ and
$\mathcal{R}$ are gauge-invariant quantities, the curvature perturbation,
characterized by the spatial Ricci scalar, is not \cite{kinney}. Hence, when a
physical interpretation of these mathematical entities is needed, one must
choose a specific gauge: the Bardeen potential represents the curvature
perturbation in the longitudinal gauge, the Mukhanov-Sasaki variable represents
the inflaton perturbation in a spatially flat gauge, and the quantity
$\mathcal{R}$ represents the curvature perturbation in the comoving gauge.

In summary, our conceptual point of view regarding the collapse states the
following: if there is no quantum collapse ($\lambda_k =0$), then $\Phi_{\nk} =
\Phi_{\nk}^R + i \Phi_{\nk}^I  \propto \bra 0 | \hat{p}_{\nk}^R |0 \ket + i
\bra
0 | \hat{p}_{\nk}^I |0 \ket=0$, with the uncertainty in $\hat{\Phi}_{\nk}^{R,I}
$ decreasing as $(-k\eta)^{-2\nu_s+1}$, and therefore the spacetime is
perfectly homogeneous and isotropic. It is only by introducing a self-induced
collapse, provided in this case by the CSL mechanism, that $\lambda_k \neq 0$,
then $\Phi = \Phi_{\nk}^R + i \Phi_{\nk}^I \propto \bra  \hat{p}_{\nk}^R \ket +
i \bra \hat{p}_{\nk}^I \ket \neq 0$ and the uncertainty of the curvature
perturbation operator decreases as $\Delta^2 \hat{\Phi}_{\nk}^{R,I} (\eta)
\simeq (-k\eta)^{-2\nu_s+1}/[\zeta_k \cos(\pi \theta_k) ] $. Also, note that
since we are treating the self-induced collapse as a sort of 'measurement
process', it must act on the hermitian operators, say $\hat \Phi_{\nk}^R$
and $\hat \Phi_{\nk}^I$. Thus, the uncertainties that are important from the
'measurement like process' are those corresponding to $\Delta^2
\hat{\Phi}_{\nk}^{R}$ and $\Delta^2 \hat{\Phi}_{\nk}^{I}$.

\subsection{Comparison to previous works}

As we have mentioned in the Introduction, the CSL model has been applied to the
inflationary universe in previous works by several authors, and they have
reached different conclusions with each other. Here we will present a brief
summary of their results and a comparison to our findings.

The first work we will address is the one presented in \cite{jmartin}. There,
the authors applied the CSL model to inflation using the Mukhanov-Sasaki variable
as the collapse operator, and assumed that the CSL parameter $\lambda$ to be the
same for all modes, i.e. independent of $k$. Those assumptions led to a scalar
power spectrum with two branches, one that is scale-invariant, and another one
with a spectral index $n_s=4$ in an evident conflict with the CMB data. In
order to suppress the conflicting branch, the CSL parameter had to be severely
constrained. Furthermore, the fact that $\lambda$ is independent of $k$ fails
to incorporate the amplification mechanism that is crucial in the CSL model. In
some sense, if $\lambda$ is the same for all modes, then small and large-scale
modes exhibit a classical behavior. Another issue mentioned by the authors is
that the uncertainty in the Mukhanov-Sasaki variable increases even if such
field variable corresponds to the CSL collapse operator.

In order to address some of the issues of \cite{jmartin}, other authors
proposed that the CSL parameter is of the form $\lambda(k,\eta) =
\lambda_0(k/k_0)^{\beta}/(-k\eta)^{\alpha}$ (with $k_0$ a pivot scale), i.e. a
function of both, the conformal time and $k$, while retaining the
Mukhanov-Sasaki variable as the collapse operator \cite{hinduesS,hinduesT}.
Hence, since the CSL parameter depends on $k$, the amplification mechanism of the CSL
model is
recovered. On the other hand, their model introduced three free parameters:
$\lambda_0, \alpha$ and $\beta$. However, by taking into account the tensor
modes \cite{hinduesT}, the modification to the scalar and tensor spectral
indices could be captured in just one effective free parameter $\delta =
3+\alpha-\beta$ and $\lambda_0$. In fact, if $\delta=0$, then their model predicts a
scale-invariant scalar power spectrum. Furthermore, if $1 < \alpha < 2$, then,
as inflation goes on, the squeezing occurs in the Mukhanov-Sasaki variable
rather than its conjugated momentum. This is because, in their approach, the
CSL parameter depends on the conformal time. Nevertheless, this dependence on
$\eta$ is inherited in their final prediction for the scalar and tensor power spectra;
thus, they choose to evaluate their power spectra at the end of inflation. This
choice, in turn, translates into an extremely small value for the CSL
parameter $\lambda_0 \simeq e^{-120}$, however, a higher value can be
achieved by bringing down the energy scale of inflation.

References \cite{hinduesS,hinduesT} can be considered as an improved version
of \cite{jmartin}; nevertheless, they share some features. First, in both approaches, if
$\lambda=0$, which means no self-induced collapse, their prediction for the
(scalar) power spectrum is exactly the same as in the standard
approach.\footnote{Actually, in the result presented in \cite{hinduesS,hinduesT}
if $\lambda_0=0$ one gets a divergent quantity for the power spectra;
nevertheless,
we think this is because of the truncated terms in their series expansions and
not because an actual problem with their model.} This contrasts with the result
presented in this letter, since in our approach, if $\lambda_k=0$ there
are no curvature perturbations at all. As a consequence, $\mP_s=0=\mP_t$. This
difference ultimately boils down to the quantum objects that are going to be
identified with the power spectrum. In the framework of
\cite{jmartin,hinduesS}, the
power spectrum corresponds exactly to the Fourier transform of the two-point
quantum correlation function, that is, $\mP_s (k) \simeq \bra \Psi | \hat
v_{\nk} \hat v_{\nk'} |\Psi \ket$, where the state $|\Psi \ket$ corresponds to
the post-collapse state, or to the Bunch-Davies vacuum if $\lambda=0$. In our
approach, we have assumed that, once the collapse took place, it can be
assumed that $\Phi_{\nk} = \bra \Psi | \hat \Phi_{\nk} | \Psi \ket $.
Therefore,
the quantum collapse is the mechanism by which the curvature perturbation
emerges. After the curvature perturbation was generated, as a result of a
stochastic process, one can calculate the power spectrum associated to a
classical stochastic field, i.e. $\mP_s (k) = \overline{\Phi_{\nk}
\Phi_{\nk'}^*} = \overline{\bra \Psi | \hat \Phi_{\nk} | \Psi \ket \bra \Psi |
\hat \Phi_{\nk'}^* | \Psi \ket }$. Another aspect that the authors in
\cite{jmartin,hinduesS} share is that they treat the field variable $v$ as their
collapse operator and then argue why it would be a desirable feature to obtain a
squeezing in that variable. From our point of view, within the cosmological
context, the choice of the collapse operator as the field variable is equally
valid as the choice of the momentum operator as we have done in the present
paper. In fact, we motivated this choice by Eq. \eqref{igualdadchingona2}, but
we could have chosen to work in the comoving gauge, in which $\mathcal{R} =
v/z$
represents the curvature perturbation, and then have argued to set $v$ as the
collapse operator. We think that in the absence of a complete relativistic
version of the CSL mechanism, it is not entirely clear which operator plays the
role of the collapse operator. On the other hand, we also think that the
squeezing of a variable does not necessarily denote that the system has become
classical. For instance, squeezed states are regarded as highly non-classical
states in Quantum Optics \cite{schumaker1,schumaker2}. Therefore, in our approach
neither the squeezing of the momentum nor the squeezing of the field variable are crucial
for the
classicalization of the perturbations. Moreover,
as we have argued, if the squeezing occurs in the momentum variable, which is
also our choice for the collapse operator, then by \eqref{igualdadchingona2},
the  ``squeezing'' also occurs in the curvature perturbation operator (because
the longitudinal gauge was selected). Additionally, \eqref{igualdadchingona2}
allows us to establish the relation $\mP_s(k) \propto \overline{\bra \Psi |
\hat p_{\nk} | \Psi \ket \bra \Psi | \hat p_{\nk'}^* | \Psi \ket } $. Likewise,
here the tensor modes are directly related to the field variable through
\eqref{igualdadchingona2h}, and in this case the collapse operator is still the
momentum operator. Thus, even if the uncertainty in $\hat v_{\nk}^{R,I}$
increases, the uncertainty in $\hat h_{\nk}^{R,I}$ (which is associated to the
tensor curvature perturbation), does not become larger. This is because of the
relation $\Delta^2 \hat h_{\nk}^{R,I} \simeq \Delta^2 \hat v_{\nk}^{R,I}/a^2$.
In both the scalar and the tensor cases, we have focused on the uncertainties
of the quantum operators that are directly related to the observables of the
CMB; namely, the operators associated to the curvature perturbations.

Finally, it is also worthwhile to mention that the conceptual point of view
adopted in this paper, regarding the role of the collapse during inflation, was
the same as the one followed in \cite{pedro}. However, in such a work, the
authors applied the CSL model to inflation within the semiclassical gravity framework,
and one important difference with our work is that, within their approach, the
amplitude of the tensor modes is exactly zero at first order in the
perturbations. Additionally, the authors in \cite{pedro} did not consider a
full quasi-de Sitter background spacetime and therefore their final expression for
the scalar power spectrum does not contain a specific prediction for the scalar
spectral index.

\section{Conclusions}
\label{conclusions}

In this work, we have adopted an important role for the self-induced collapse
of the wave function characterized by the CSL model, i.e. the collapse acts as the
main agent for the emergence of the primordial curvature perturbations. We are
aware that other people in the scientific community might prefer to adopt the Everett ``many-world''
interpretation of quantum theory plus decoherence in the case of the inflationary
universe. However, we have presented that a dynamical reduction of the wave function,
provided by the CSL model, can be a viable alternative explanation in the sense that is
consistent with observations and also addresses the standard ``measurement problem'' of
quantum theory.

Specifically, by focusing on the curvature perturbation operators, we have presented predictions
for the scalar and tensor power spectra \eqref{psfinal} that are quite
consistent with the standard prediction. In particular, the scalar and tensor
spectral indices are exactly the same, as well as the scalar $A_s$ and tensor
$A_t$ amplitudes. The tensor-to-scalar ratio is also of the same form as in the
traditional approach. On the other hand, the addition of the CSL mechanism is
reflected in the appearance of the function $F(\lambda_k,\nu)$ [defined in
\eqref{F}] multiplying the standard prediction for the scalar and tensor power spectra.
However, if $\nu \simeq 1/2$ and the CSL parameter is of the form
$\lambda_k = \lambda_0/k$, then $F(\lambda_k,\nu)$ becomes practically
independent of $k$, and we recover the standard prediction for the power
spectra. This allows us to give a roughly estimate for the CSL parameter
$\lambda_0 \simeq 10^{-21}$ s$^{-1}$ (similar to the one suggested by the GRW model,
$\lambda_{\textrm{GRW}} = 10^{-16}$ s$^{-1}$), which can be refined by using the full
CMB data set, what is left for a future work. On the other hand, small variations
of the recipe $\lambda_k = \lambda_0/k$ will result in different predictions
from the standard case and that could be confronted with the observational data.

Our prediction for the power spectra differs from the ones obtained in previous
works \cite{jmartin,hinduesS,hinduesT} in which the CSL mechanism was also
applied to the inflationary universe. In particular, in contrast to Ref. \cite{jmartin},
we
have captured the amplification mechanism by assuming a $k$ dependence of the CSL
parameter. Additionally, our final result for the power spectra is not time
dependent and we have only one free parameter $\lambda_0$ to be adjusted using
the observational data. This is different from the findings of Ref.
\cite{hinduesS}, in which their final predictions for the power spectra depend on the
conformal
time and their model introduces more than one free parameter. The reasons for these
differences are traced back to the conceptual point of view regarding the
collapse of the wave function during inflation. Specifically, if $\lambda_k=0$,
within the framework of Refs. \cite{jmartin,hinduesS}, then their model
recovers the standard prediction. On the other hand, in our approach, if $\lambda_k=0$, there is
no quantum collapse. Then the tensor and scalar power spectra are exactly zero,
which means that the spacetime does not contain any perturbations at all.
Another important (conceptual) difference from previous research is that the
squeezing of the field and momentum variables seems to play a relevant role in the
approach
of Refs. \cite{jmartin,hinduesS,hinduesT}.  In our approach, we have shown that the
uncertainty of the
curvature perturbation operator (either in the comoving or in the longitudinal gauge)
decreases as inflation goes on;
however, the self-induced collapse is still needed for generating the primordial
perturbation.

Our result, unlike the case of semiclassical gravity, predicts non-null tensor
modes at first order in the perturbations, and it is also consistent with the
findings of \cite{mauro}, where the tensor spectrum and the tensor-to-scalar
ratio were calculated within a generic collapse scheme.

We conclude that, from the phenomenological point of view, the CSL mechanism can
successfully be used to improve the standard inflationary paradigm by explaining
the origin of the primordial perturbations and driving the quantum-to-classical
transition.
However, there are still some aspects concerning the CSL model that need to be
addressed further. For example, it is well known that the CSL model violates
energy conservation, which can lead to divergences in the energy-momentum
tensor. Also, the CSL model is nonrelativistic, hence, the choice of the
collapse operator and the $k$ dependence in the $\lambda_k$ parameter is at this
point purely phenomenological. Therefore, we think that the early universe
provides a natural laboratory to test new ideas that would allow us to reach a
deep understanding of Nature.

\begin{acknowledgements}
G.R.B. is supported by CONICET (Argentina). G.R.B. acknowledges
support from the PIP 112-2012-0100540 of CONICET (Argentina). G.L.'s research was
funded by Consejo Nacional de Investigaciones Cient\'{i}ficas y T\'{e}cnicas,
CONICET (Argentina) and by Consejo Nacional de Ciencia y Tecnolog\'ia, CONACYT (Mexico).
\end{acknowledgements}

\appendix

\section{Homogeneity and isotropy of the vacuum state}\label{appA}

The vacuum state is
defined by $\hat{a}_{\nk} |0 \ket =0$, which  is the state of the quantum field after a
few $e$-folds of inflation; usually such a vacuum state is chosen  to be the Bunch-Davies
vacuum. Moreover, the operator generating spatial translations is
$\hat P_j = \sum_{\nk} k_j \hat{a}_{\nk}^{\dag} \hat{a}_{\nk} $ with $j=1,2,3$. Therefore,
applying a spatial translation to the vacuum, given by the displacement vector
$\textbf{S}$,  yields $\exp[i S_j \hat{P}_j] |0 \ket = |0 \ket$, thus, the vacuum state is
homogeneous. One can similarly check the isotropy of the vacuum state by considering the
behavior of the state under rotations. Consequently, the state of the field is homogeneous
and isotropic.

During the inflationary regime, the quantum field is usually represented by the
Mukhanov-Sasaki field variable $\hat v$ [see Eq. \eqref{MS}]. One can easily show that,
for a mode $\hat{v}_{\nk}$ of the quantum field, one has $\bra 0 | \hat{v}_{\nk} | 0\ket =
0$ and $\bra 0 | \hat{v}_{\nk}^2 | 0\ket \neq 0$, hence, the quantum uncertainty
$\Delta^2 \hat{v}_{\nk} \equiv \bra 0 | \hat{v}_{\nk}^2 | 0\ket - \bra 0 | \hat{v}_{\nk} |
0\ket^2$ is clearly not zero.

Thereupon,  the fact that the quantum uncertainty is not zero (or that the
system contains vacuum fluctuations) does not mean that the system contains
inhomogeneities of any definite size. As we have shown, the quantum state
characterizing each mode of the field $\hat{v}_{\nk}$ is homogeneous and isotropic and
the symmetries of the quantum state encode the symmetries of the physical system.

\section{Possible ways to relate $\hat \Phi$ and $\Phi$}\label{appB}

In this appendix we will broaden the discussion to other possible ways to relate the
quantum operator associated to the curvature perturbation $\hat \Phi$ and its
classical stochastic value $\Phi$ within the CSL framework.

As we mentioned in Sect. \ref{cslinflation}, for a Fourier mode, the usual way to relate
$\hat \Phi_{\nk}$ and $\Phi_{\nk}$ is through $\Phi_{\nk}
=  \sqrt{\bra 0 | \hat \Phi_{\nk}^2 |0\ket}e^{i \alpha_{\nk}}$ with $\alpha_{\nk}$ a
random phase (the relation is valid only when the mode has become super-horizon).
Note that since $\bra 0 | \hat \Phi_{\nk} |0\ket =0$, the quantum uncertainty
$\Delta^2 \hat{\Phi}_{\nk}$ is exactly equal to   $\bra 0 | \hat \Phi_{\nk}^2 |0\ket$.

Although the works in Refs. \cite{jmartin,hinduesS,hinduesT} differ in some ways of
implementing the CSL model to the inflationary universe, they do share the same approach
in relating $\hat \Phi_{\nk}$ and $\Phi_{\nk}$. That is, they both use the relation

\beq\label{1app}
\Phi_{\nk} =  \left(\Delta^2 \hat{\Phi}_{\nk}\right)^{1/2}_{\textrm{CSL}}
e^{i \alpha_{\nk}}
\eeq
where $(\Delta^2 \hat{\Phi}_{\nk})^{1/2}_{\textrm{CSL}} $ represents the
square root of the quantum uncertainty associated to the operator $\hat \Phi_{\nk}$
evaluated in the state that results from the non-unitary evolution given by the CSL
mechanism, explicitly

\beq
(\Delta^2 \hat{\Phi}_{\nk})_{\textrm{CSL}} = \bra \Theta |
\hat{\Phi}_{\nk}^2 | \Theta \ket - \bra \Theta | \hat{\Phi}_{\nk} | \Theta \ket^2.
\eeq

It is thus clear that if there is no CSL evolution, i.e. if $|\Theta \ket = |0\ket$ then
the standard way of relating $\hat \Phi_{\nk}$ and $\Phi_{\nk}$ coincides with the one
provided by Eq. \eqref{1app}. That can explain the fact that if one considers the CSL
parameter $\lambda_{k}=0$ in the expressions
for the scalar power spectrum in Refs. \cite{jmartin,hinduesS,hinduesT},  i.e. the CSL
evolution is ``switched off,'' then the prediction for the scalar power spectrum is
exactly the same as the standard one. Moreover, that could also explain why in both works
the CSL parameter $\lambda_{k}$ has to be extremely small in order to fit the
predictions for the scalar power spectrum with the observational data. That is, if
$\lambda_k \to 0$, then $|\Theta \ket \to |0\ket$ and both predictions for the
scalar power spectrum (the traditional and the CSL one of Refs.
\cite{jmartin,hinduesS,hinduesT}), are the same. Since we know that the traditional
prediction for the scalar spectrum fits very well the observational data, it is
plausible  to think that if $\lambda_{k} << 1$, then the scalar spectrum of Refs.
\cite{jmartin,hinduesS,hinduesT} will be consistent with the observations too.

\end{document}